\documentclass[12pt]{iopart}
\usepackage{iopams}
\usepackage{graphicx}
\newcommand{\C}{\mathbb C}

\newcommand{\I}{\mathbb I}

\newcommand{\N}{\mathbb N}

\newcommand{\R}{\mathbb R}

\def\lg{\langle }
\def\rg{\rangle }

\def\deq{\stackrel{\mathrm{def}}{=}}

\begin{document}
\title[ Complex and real Hermite polynomials and related quantizations]{ Complex and real Hermite polynomials and related quantizations}

\author{Nicolae Cotfas$^1$, Jean Pierre Gazeau$^2$, and Katarzyna G\'{o}rska$^3$}
\address{$^1$Faculty of Physics, University of Bucharest, PO Box 76 - 54, Post Office 76, Bucharest, Romania}
\address{$^2$Laboratoire APC, Universit\'e Paris 7-Denis Diderot, 10, rue A. Domon et L. Duquet, 75205 Paris Cedex13, France}
\address{$^3$Institute of Physics,
Nicolaus Copernicus University,
ul. Grudziadzka 5/7,
87-100 Torun, Poland}
\eads{\mailto{ncotfas@yahoo.com}, \mailto{gazeau@apc.univ-paris7.fr}, \mailto{dede@fizyka.umk.pl}}
\begin{abstract} 
It is known that the anti-Wick (or standard coherent state) quantization of the complex plane produces both canonical commutation rule and  quantum spectrum of the harmonic oscillator (up to the addition of a constant). In the present work, we show that these two issues are not necessarily coupled: there exists a  family of separable Hilbert spaces, including the usual Fock-Bargmann space, and in each element in this family there exists an overcomplete set of unit-norm states resolving the unity. With the exception of the Fock-Bargmann case, they all produce non-canonical commutation relation whereas  the  quantum spectrum of the harmonic oscillator remains the same up to the addition of a constant. The statistical aspects of these non-equivalent coherent states quantizations are investigated. We also explore the localization aspects in the real line yielded by similar quantizations based on real Hermite polynomials. 
\end{abstract}
\maketitle
\section{Introduction}
\label{secintro}

It is well known that the anti-Wick (or Klauder-Berezin-Toeplitz) quantization (see for instance \cite{alienglis} and references therein) of the complex plane equipped with the Lebesgue measure yields both canonical commutation rule, $[\hat q, \hat p]=i\hbar \, \I$,  and  quantum spectrum of the harmonic oscillator (namely $\hbar \omega(\hat N + 1/2)$ up to the addition of $1/2$). The aim of  this paper is to prove that these two issues are not necessarily coupled: there exists a  discrete family of separable Hilbert subspaces  $\mathcal{K}_s$, $s\in \N$,  in $L^2( \C, d^2z/\pi)$, including the ``canonical'' subspace Fock-Bargmann, and in each element in this family there exists an overcomplete set of states resolving the unity and producing, with the exception of the Fock-Bargmann case,  \underline{non}-canonical commutation relation  \underline{and} the same quantum spectrum of the harmonic oscillator up to the addition of the constant $2s + 1/2$. Each $\mathcal{K}_s$ is the closure of the linear span of complex Hermite polynomials \cite{alibaga, ghanmi} weighted by a Gaussian, $e^{-\vert z\vert^{2}/2}\, h^{s+n, s}(z, \bar z)$.

The organization of the paper is as follows. In Section \ref{secwick} we recall some well-known facts about the anti-Wick or standard coherent state quantization and make  comparison with the canonical quantization. Then, in Section \ref{seccsquant}, we present a general construction of coherent states (CS) and we describe the corresponding CS quantization. In the following sections, the procedure  is worked out with coherent states based on  complex and real Hermite polynomials. The complex Hermite polynomials are defined in Section \ref{seccomherm} and the corresponding quantization of the complex plane is  implemented in Section \ref{sechermquant}. Its remarkable feature is the appearance of  a new commutation rule for the lowering and raising operators, and so for the position and momentum operator, where is involved an extra term proportional to the projector on the ground state. Notwithstanding, we obtain for the energy spectrum of the CS quantized harmonic oscillator the same as for the usual one up to addition of a constant defined by the class of considered complex Hermite polynomials. We examine in Section \ref{susy} a possible connection of our results with supersymmetric quantum mechanics.    Some statistical aspects of the complex Hermite polynomial  coherent states and the corresponding quantization are examined in Section \ref{secstat}. In the same vein, we explore in Section \ref{secrealmerm} the  quantization of the real line with coherent states in finite dimensional Hilbert spaces constructed with real Hermite polynomials and we study the resulting localization properties. It turns out that for a given dimension the position operator is the same as the position operator derived from  the corresponding finite dimensional approximation of the usual  quantum mechanics. We give in  Section \ref{seconc} some indications for future developments issued from our work.

\section{Anti-Wick or coherent state versus canonical quantization}
\label{secwick}

The  anti-Wick quantization, to which we prefer the name of coherent state (CS) quantization,  consists in
starting from the plane $\R^2 \simeq \C = \{  z = \frac{1}{\sqrt2}(q+ip) \}$, where we put $\hbar
=1$ for convenience, equipped  with its Lebesgue measure $\mu(dz\,d\bar z) \equiv \frac{1}{\pi}\,  d^2 z$ with $d^2z =d\Re z\,d\Im z$, and viewed as the  phase space for the motion of a particle on the line. In the Hilbert space 
$L^2( \C, \mu(dz\,d\bar z))$ of all  complex-valued functions on the complex plane which are square-integrable with respect to this  measure, we choose the orthonormal set 
 formed of the normalized
powers of the conjugate of the complex variable $z$ weighted by the Gaussian , \emph{i.e.} $\phi_n (z) \equiv e^{-\vert z \vert^2/2}\,\frac{{\bar z}^n}{\sqrt{n!}}$ with $n \in \N$.  This set is an orthonormal basis for the so-called Fock-Bargmann Hilbert subspace, here denoted by $\mathcal{K}_0$,  in $L^2( \C, \mu(dz\,d\bar z))$. Let ${\mathcal H}$ be a separable Hilbert space (e.g. a Fock space) with orthonormal basis $\{|e_n\rg \, , \, n \in \N\}$ (e.g. the ``number states'' $|n\rg$).  We then consider  the following infinite linear superposition in ${\mathcal H}$
\begin{equation}| z\rangle =   \sum_n  \bar \phi_n (z) | n\rangle =e^{-\frac{\vert z \vert^2}{2}} \sum_{n\in \N}  \frac{z^n}{\sqrt{n!}}| e_n\rangle,
\label{scs}
\end{equation}
They are  the well-known Schr\"odinger-Klauder-Glauber-Sudarshan, or simply standard, coherent states. From the numerous properties of these states \cite{K1963b,G2009} we retain here two features, namely normalization and unity resolution:
\begin{equation}\langle z\, | z \rangle = 1\, ,   \qquad 
\frac{1}{\pi}\int_{\C}  | z\rangle \langle z| \, d^2 z= \I_{{\mathcal H}}.
\label{pscs}
\end{equation}
CS quantization means that  a classical observable $f$, that is a (usually supposed smooth) function of phase space variables $(q, p)$ or equivalently of $(z,\bar z)$, is transformed through the operator integral
\begin{equation}
\frac{1}{\pi}\int_{\C} f( z, \bar z)\, | z\rangle \langle z| \, d^2 z = A_f,
\label{quantizer}
\end{equation}
into an operator $A_f$ acting on the Hilbert space ${\mathcal H}$.  We  get for the most basic one,
\begin{equation}\frac{1}{\pi}\int_{\C}  z\, | z\rangle \langle z| \, d^2 z = \sum_n \sqrt{n+1} 
| n\rangle \langle n+1| \equiv a,
\label{low}
\end{equation}
which is the lowering operator, $a | e_n\rangle = \sqrt{n} | e_{n - 1}\rangle$. We easily check that the coherent states are eigenvectors of $a$ : $a |z\rangle = z |z\rangle$. The  adjoint
$a^{\dagger}$ is obtained by replacing $z$ by $\bar z$ in (\ref{low}), and we get the
factorisation $\hat N = a^{\dagger}a$ for the number operator, $\hat N |e_n\rangle = n |e_n\rangle$, together with the  commutation rule $\lbrack a, a^{\dagger} \rbrack = \I_{{\mathcal H}}$.  The  lower symbol or expected value of the number operator  $\langle z | \hat N |z \rangle $ is precisely $\vert z \vert^2$.  
From $q = \frac{1}{\sqrt{2}}(z + \bar z)$ and  $p = \frac{1}{\sqrt{2}i}(z - \bar z)$, one easily infers by linearity that the canonical position $q$ and momentum  $p$
map to the quantum observables $\frac{1}{\sqrt{2}}(a + a^{\dagger}) \equiv Q$ and $\frac{1}{\sqrt{2}i}(a - a^{\dagger})
\equiv P$ respectively. In consequence,  the self-adjoint operators $Q$ and $P$ obey the canonical
commutation rule $\lbrack Q, P \rbrack = i\I_{{\mathcal H}}$, and for this reason fully deserve  the name of position and momentum operators of the usual (galilean) quantum mechanics, together with all  localisation properties specific to the latter.  Let us now CS quantize the classical harmonic oscillator Hamiltonian
$H = \frac{1}{2} (p^2 + q^2)= \vert z \vert^2$:
\begin{equation}
\label{modzsq}
A_{H} = A_{\vert z \vert^2}=   \hat N +  \I_{{\mathcal H}} \,.
\end{equation}
We see with this elementary example that the  CS quantization does not fit exactly with the ``canonical'' one, which consists in just replacing $q$ by $Q$ and $p$ by $P$ in the  expressions of the observables $f(q,p)$ and next proceeding with a symmetrization in order to comply with self-adjointness. In fact, the quantum Hamiltonian obtained 
through this usual ansatz  is equal to $\hat{H} = \frac{1}{2} (P^2 + Q^2)=\hat N + (1/2)\, \I_{{\mathcal H}} $. In the present case, there is a shift by  $1/2$ between the  spectrum of  $\hat{H}$ and the CS quantized Hamiltonian $A_H$. Actually, no physical experiment can discriminate between those two spectra that differ from each other by a simple shift (for a thorough discussion on this point, see for instance \cite{kastrup}). 

\section{Coherent state quantization: the general setting}
\label{seccsquant}
Let $\Sigma$ be a set of parameters equipped with a measure $\mu$ and its associated Hilbert space $L^2(\Sigma, \mu)$ of complex-valued square integrable functions with respect to $\mu$. Let us choose 
in $L^2(\Sigma, \mu)$ a finite or countable  orthonormal set $\mathcal{O}=\{\phi_n\, , \, n = 0, 1, \dots \}$:
\begin{equation}\label{eqI1}
\lg \phi_m | \phi_n \rg = \int_{\Sigma}\overline{\phi_m(\alpha)}\, \phi_n(\alpha)\, \mu(d\alpha) = \delta_{mn}\, .
\end{equation}
In case of infinite countability, this set must obey the (crucial) finiteness condition:
\begin{equation}\label{eqI2}
\sum_{n} \vert \phi_n (\alpha)\vert^2 \deq \mathcal{N}(\alpha) < \infty \,  \quad \mathrm{a.e.}\, . 
\end{equation}
Let $\mathcal{H}$ be a separable complex Hilbert space with orthonormal basis $\{|e_n\rg\, , \, n = 0, 1, \dots \}$ in one-to-one correspondence with the elements of  $\mathcal{O}$. From Conditions (\ref{eqI1}) and (\ref{eqI2}) there results that the  family of normalized ``coherent'' states $\mathcal{F}_{\mathcal{H}}= \{|\alpha\rg\, , \, \alpha \in \Sigma \}$ in $\mathcal{H}$, which are defined by
\begin{equation}\label{eqI3}
|\alpha\rg = \frac{1}{\sqrt{\mathcal{N}(\alpha)}}\sum_n \overline{\phi_n(\alpha)}\, |e_n\rg\, ,
\end{equation}
resolves the identity in $\mathcal{H}$:
\begin{equation}\label{eqI4}
\int_\Sigma \mu(d\alpha) \,\mathcal{N}(\alpha) \, |\alpha\rg \lg \alpha | = \I_{{\mathcal H}}\, .
\end{equation}
Such a relation allows us to implement a \emph{coherent state or frame quantization} of the set of parameters $\Sigma$ by associating to a function $\Sigma \ni \alpha \mapsto f(\alpha)$ that satisfies appropriate conditions the following operator in $\mathcal{H}$:
\begin{equation}\label{eqI5}
f(\alpha) \mapsto A_f  \deq \int_\Sigma\mu(d\alpha) \,\mathcal{N}(\alpha) \, f(\alpha)\, |\alpha\rg \lg \alpha |\, .
\end{equation}
Operator $A_f$ is symmetric if $f(\alpha)$ is real-valued, and is bounded if $f(\alpha)$ is bounded. The original $f(\alpha)$ is a ``upper symbol'', usually non-unique,  for the operator $A_f$. It will be called a 
\emph{classical} observable with respect to the family $\mathcal{F}_{\mathcal{H}}$ if the so-called 
``lower symbol" $\check{A}_f (\alpha)\deq\lg \alpha | A_f | \alpha \rg$ of $A_f$ has mild functional properties to be made precise according to further topological properties granted to the original set $\Sigma$.

\section{Complex Hermite polynomials}
\label{seccomherm}
Let $r$ and $s$ be nonnegative integers. Complex Hermite polynomials are defined as \cite{alibaga, ghanmi}:
\begin{equation}\label{eqCH1} \fl
h^{r,s}(z,\bar{z}) = (-1)^{r+s}\, e^{\vert z \vert^2}\, \frac{\partial^r}{\partial z^r}\, \frac{\partial^s}{\partial \bar z^s}\, e^{-\vert z \vert^2} = \sum_{k=0}^{\min(r,s)} \frac{(-1)^k}{k!}\,\frac{r!s!}{(r-k)!(s-k)!}\, z^{s-k}\, \bar z^{r-k}\, .
\end{equation}
They form a complete orthogonal system in the Hilbert space $L^2\left(\C\,,\, e^{-\nu\vert z \vert^2} \,d^2z\right)$ with $\nu > 0$. 
Suppose now that $r \geq s$. Then the corresponding  polynomials can be written  in terms of confluent hypergeometric functions or in terms of associate Laguerre polynomials:
\begin{eqnarray}\nonumber
h^{s+n, s}(z,\bar{z}) &=& s! (s+n)!\, \bar{z}^{n}\, \sum_{k = 0}^{s} \frac{(-1)^{s-k}}{(s-k)!}\,\frac{\vert z\vert^{2k}}{k!\,(k+n)!}, \\[0.5\baselineskip] \label{eqCH2}
\nonumber&=& \frac{(-1)^{s} \,(s+n)!}{n!} \, \bar{z}^{n} \,_{1}F_{1}(-s; n + 1; \vert z\vert^{2}) \\
&=&  (-1)^{s} \,s!\, \bar{z}^{n} \, L_s^{(n)}(\vert z\vert^{2}) \, ,
\end{eqnarray}
where $r-s=n\in\mathbb{N}$. In particular, for $s=0$ and $1$, the expression  (\ref{eqCH2}) reduces, respectively,  to $\bar{z}^{n}$ and $\bar{z}^{n}(\vert z\vert^{2} - n-1)$. For a fixed $s$ we have an infinite  family of complex polynomials  of degree $n+2s$ in variables $z$ and $\bar z$, and which are pairwise orthogonal. Precisely, by using the relation (2.20.1.19) in \cite{Prud1}, we obtain:
\begin{equation}\label{eqCH3}
\frac{1}{\pi}\,\int_{\C} d^{2}z\, e^{-\vert z\vert^{2}}\, h^{s+n, s}(z, \bar z) \overline{h^{s+n', s}(z, \bar z)} = \left\{ 
\begin{array}{lll}
s!\, (s+n)! & {\rm if} & n= n'\\[2mm] 
0 & {\rm if} & n\neq n'\,.
\end{array} \right.
\end{equation}

The functions $h^{s+n,s}$ are related through the ladder operators
\begin{equation}\fl \left\{ 
\begin{array}{l}
\left( -\frac{\partial }{\partial z}+\bar z\right) h^{s+n,s}=h^{s+n+1,s} \\[5mm]
\frac{\partial }{\partial \bar z}h^{s+n+1,s}=(s\!+\!n\!+\!1)\, h^{s+n,s}
\end{array} 
\qquad \right\{
\begin{array}{l}
\left( -\frac{\partial }{\partial \bar z}+z\right) h^{s+n,s}=h^{s+n,s+1}\, ,  \\[5mm]
\frac{\partial }{\partial z}h^{s+n,s+1}=(s\!+\!1)\, h^{s+n,s}\, .
\end{array}
\end{equation}

We define in the Hilbert space  $L^2( \C, d^2z/\pi)$ the Hilbert subspace $\mathcal{K}_s$ as the closure of the linear span of the set of orthonormal functions defined as 
\begin{equation} \label{phisn}
\begin{array}{ll}
\nonumber \phi_{n;s}(z) &\deq \frac{1}{\sqrt{s! (s+n)!}}e^{-\vert z\vert^{2}/2}\, h^{s+n, s}(z, \bar z)\\
& = (-1)^{s}  \sqrt{\frac{s!}{(s+n)!}}\,e^{-\vert z\vert^{2}/2}\,\bar{z}^{n} \, L_s^{(n)}(\vert z\vert^{2})\, .  
\end{array}
\end{equation}
The functions $\phi _{n;s}$ are related through the ladder operators
\begin{equation}
\label{phisnlad}
\begin{array}{l}
\left( \frac{\partial }{\partial \bar z}+\frac{z}{2}\right)\phi _{n+1; s}=\sqrt{s+n+1}\, \phi _{n; s}\\[3mm]
\left( -\frac{\partial }{\partial z}+\frac{\bar z}{2}\right)\phi _{n; s}=\sqrt{s+n+1}\, \phi _{n+1; s}.
\end{array}
\end{equation}
 The ``canonical'' Fock-Bargmann subspace corresponds to $s=0$. 
 We thus obtain  a countably infinite family of orthogonal Hilbert subspaces $\mathcal{K}_s$.

\section{Complex Hermite polynomial quantization}
\label{sechermquant}
Following the guideline indicated in Section \ref{seccsquant}, for a fixed $s$,  we  construct the coherent states based on complex Hermite polynomials  as the  infinite linear combination of orthonormal elements $|n;s\rangle$ of some separable Hilbert space $\mathcal{H}_{s}$ 
\begin{equation}\label{eqCS0} \fl
|z;s\rangle = \frac{1}{\sqrt{e^{-\vert z \vert^2}\,\mathcal{N}_{s}(\vert z\vert^{2})}}\,\sum_{n=0}^{\infty}\,\overline{\phi_{n;s}(z)}|n;s\rangle = \frac{1}{\sqrt{\mathcal{N}_{s}(\vert z\vert^{2})}}\,\sum_{n=0}^{\infty}\,\frac{\overline{h^{s+n,s}(z,\bar{z})}}{\sqrt{s!\,(s+n)!}}\, |n;s\rangle \, , 
\end{equation}
where the normalization factor is defined as 
\begin{equation}\label{eqCS2}
\mathcal{N}_{s} (\vert z \vert^2)\,=\, \sum_{n = 0}^{\infty}\frac{\vert h^{s+n,s}(z, \bar z)\vert^{2}}{s!(s+n)!}\, .\end{equation}
Note the change of notation in regard with Eq. (\ref{eqI2}) in order to delete  the Gaussian factor. Also, we could choose all spaces $\mathcal{H}_{s}$ as identical, e.g. the Fock space spanned by number states $|n\rg$, or the Hilbert space $L^2(\R, dx)$, in which case there is no need to specify the parameter $s$. On the other hand we could choose $\mathcal{H}_{s} = \mathcal{K}_{s}$ and identify the states $|n;s\rg$ with the functions $\phi_{n;s}$. 

The series (\ref{eqCS2}) can be easily summed  for lower  values of $s$, e. g. for $s = 0$ and $1$:  they are respectively  equal to $e^{\vert z\vert^2}$ and $e^{\vert z\vert^{2}}-\vert z\vert^{2}$. If we use the definition of the complex Hermite polynomials (\ref{eqCH2}) in eq. (\ref{eqCS0}), we obtain the alternative  form: 
\begin{equation}\label{eqCS1}
|z; s\rangle \,=\, \frac{(-1)^{s}}{\sqrt{\mathcal{N}_{s}(\vert z\vert^{2})}} \, \sum_{n=0}^{\infty}\,
{\left(
\begin{array}{c}
s\!+\!n\\
s
\end{array}
\right)}^{-1/2}\, 
\frac{z^{n}}{\sqrt{n!}} \,L_s^{(n)}(\vert z\vert^{2}) \, |n;s\rangle\, , 
\end{equation}
and for the normalization function,
\begin{equation}\label{eqCS3}
\mathcal{N}_{s} (\vert z \vert^2)\,=\, \sum_{n = 0}^{\infty} \frac{s!}{ (s+n)!} \, \vert z\vert^{2n}\,\left(L_s^{(n)}(\vert z\vert^{2}) \right)^2\, .
\end{equation}
With the help of this form it can be easily checked that for $s=0$ they are the standard coherent states, but for the remaining values of $s$ we are in presence of  some deformation of the standard $|z; 0\rangle \equiv |z\rg$. Therefore we have with Eq. (\ref{eqCS1})  an infinite family of coherent states families, which is labeled by $s\in \N$. 

We next proceed with the corresponding coherent state quantization, starting as usual with the simplest functions $f(z, \bar{z}) = z$ and $\bar{z}$. With the help of eqs. (\ref{eqI5}) and (2.20), (2.19.23.6) in \cite{Prud1}, we get
\begin{equation}\label{eqCS3} 
\begin{array}{l}
A_{z} \,=\, \sum_{n=0}^{\infty}\, \sqrt{s+n+1}\, |n;s\rangle\langle n+1; s|,\\[3mm]
A_{\bar{z}} \,=\, \sum_{n=0}^{\infty}\, \sqrt{s+n+1}\, |n+1; s \rangle \langle n;s|.
\end{array}
\end{equation}
The lowering $A_{z}$ and uppering $A_{\bar{z}}$ fulfill a new commutation relation 
\begin{eqnarray}\nonumber
\left[A_{z}, A_{\bar{z}}\right] &=& \sum_{n=0}^{\infty}\, (n+s+1) \, \left(|n;s\rangle\langle n;s| - |n+1;s\rangle\langle n+1; s|\right), \\
&=& \mathbb{I}_{\mathcal{H}_{s}} + s |0;s\rangle\langle 0;s|. \label{eqCS4}
\end{eqnarray}
 The equation (\ref{eqCS4}) for $s=0$ leads to the usual commutation rule for $A_{z}$, $A_{\bar{z}}$, this is, $[A_{z}, A_{\bar{z}}] = \mathbb{I}_{\mathcal{H}_{0}}$. In the case of other value of $s$, there is an extra term proportional to the orthogonal projector on the ``ground state''  $|0;s\rangle$.

The position $\hat{q}$ and momentum $\hat{p}$ operators are  easily obtained by using the quantized version of the relations $q = (z + \bar{z})/\sqrt{2}$, $p = -i(z - \bar{z})/\sqrt{2}$, where the coordinates $q$, $p$ and $z$, $\bar{z}$ are replaced by operators $\hat{q}$, $\hat{p}$ and $A_{z}$, $A_{\bar{z}}$. Now, with the help of eqs. (\ref{eqCS3}) we have
\begin{equation}\label{eqCS5} 
\hat{q} \,=\, \sum_{n=0}^{\infty} \sqrt{\frac{s+n+1}{2}}\, \left(|n;s\rangle\langle n+1;s| + |n +1;s\rangle\langle n;s|\right)
\end{equation}
\begin{equation}\label{eqCS6} 
\hat{p} \,=\, -i\sum_{n=0}^{\infty} \sqrt{\frac{s+n+1}{2}}\, \left(|n;s\rangle\langle n + 1;s| - |n+1; s\rangle\langle n;s|\right)
\end{equation}
In explicit matrix form we have for $\hat{q} $
\begin{equation}
\hat{q} = \left(
\begin{array}{llllll}
	0 & \sqrt{\frac{s+1}{2}} & 0 & \cdots \\
	\sqrt{\frac{s+1}{2}} & 0 & \sqrt{\frac{s+2}{2}} & \cdots \\
	0 & \sqrt{\frac{s+2}{2}} & 0 & \ddots  \\
	\vdots &  & \ddots & \ddots \\
\end{array}\right)\, ,
\end{equation}
and a similar expression for $\hat{p}$. Their commutation rule, $[\hat{q}, \hat{p}] = i[A_{z}, A_{\bar{z}}]$, is ``almost'' canonical, in the sense that  like for (\ref{eqCS4}) there is the  extra projector on the ground state multiplied by  $i\,s$. 

We now turn our attention to the energy operator for the one-dimensional quantum harmonic oscillator. There are at least  two expressions for it. The first one which appears to us as the most natural is issued from the CS quantization of   the classical Hamiltonian, 
$H \,=\, (q^{2} + p^{2})/2 = \vert z \vert^2$. Its quantum version  $A_{\vert z\vert^{2}, s}$ is easily calculated and reads as the diagonal operator
\begin{equation}\label{eqCS7}
A_{\vert z\vert^{2}, s} \,=\, \sum_{n=0}^{\infty}\, (n+ 2s + 1)\, |n;s\rangle\langle n;s|\, .
\end{equation}
This entails  that the lowest state $|0; s\rangle$ has energy  $(2s + 1)$ and  that the energy levels are equidistant by 1, like for the energy levels of the canonical case. 

The alternative to this direct CS quantization is to use the standard ansatz which consists in replacing $q$ by $\hat q$ and $p$ by $\hat p$ in the expression of the classical observable $H \,=\, (q^{2} + p^{2})/2$. This leads to the operator $\hat{H} = (\hat{q}^{2} + \hat{p}^{2})/2$, where, respectively, $\hat{q}$ and $\hat{p}$ are given by (\ref{eqCS5}) and (\ref{eqCS6}). Now, we get
\begin{equation}\label{eqCS8}
\hat{H}\!=\! \frac{s\!+\!1}{2}\, |0;s\rangle\langle 0;s| + \sum_{n\geq 1}\, (n \!+\! s\!+\! 1/2)\, |n;s\rangle\langle n;s|.
\end{equation}
The distance between the  first and second level is $s/2 + 1$, whereas the distance between the upper levels (e. g., third and  second level and so on) is constant and equal to 1. It is obvious that for $s=0$ eqs. (\ref{eqCS7}) and (\ref{eqCS8}) are the same. The distinctions between them hold for  $s\geq 1$, for which there is  a shift of the ground state energy. 

It is interesting to examine  the respective lower symbols of $\hat q$ and $\hat p$.
\begin{equation}
\check{q} \,=\, \langle z;s|\hat{q}|z;s\rangle, \quad \check{p} \,=\, \langle z;s|\hat{p}|z;s\rangle.
\end{equation}
We get
\begin{equation} \fl
\check{q} \,=\, \frac{q}{\mathcal{N}_s(\vert z\vert^{2})} \,\sum_{n=0}^{\infty} 
\left(
\begin{array}{c}
s\!+\!n\!+\!1\\
s
\end{array}
\right)
\frac{\vert z\vert^{2n}}{n!} \,_{1}F_{1}(-s; n+1; \vert z\vert^{2}) \,_{1}F_{1}(-s; n+2; \vert z\vert^{2})
\end{equation}
\begin{equation} \fl
\check{p} \,=\frac{p}{\mathcal{N}_s(\vert z\vert^{2})} \,\sum_{n=0}^{\infty} 
\left(
\begin{array}{c}
s\!+\!n\!+\!1\\
s
\end{array}
\right)
\frac{\vert z\vert^{2n}}{n!} \,_{1}F_{1}(-s; n+1; \vert z\vert^{2}) \,_{1}F_{1}(-s; n+2; \vert z\vert^{2})
\end{equation}
In the simplest cases $s \,=\, 0$ and  $s\,=\, 1$ we obtain respectively
\begin{equation}
\check{q} \, = q\, , \qquad \check{p} \, = p\, ,
\end{equation}
\begin{equation}
\check{q} \,=\, q\, \left(1 + \frac{1}{e^{\vert z\vert^{2}} - \vert z\vert^{2}}\right), \qquad \check{p} \,=\, p\, \left(1 + \frac{1}{e^{\vert z\vert^{2}} - \vert z\vert^{2}}\right),
\end{equation}
The first case $s=0$ yields a well-known result, whilst the second case displays an interesting deformation of the complex plane essentially concentrated around the origin.

\section{A possible interpretation in terms of Supersymmetric Quantum Mechanics (SUSYQM)} 
\label{susy}
It is well-known that the harmonic oscillator Hamiltonian
\begin{equation}
{\bf H}=-\frac{1}{2}\, \frac{d^2}{dx^2}+\frac{1}{2}x^2
\end{equation}
has the eigenvalue spectrum
\begin{equation}
\frac{1}{2},\quad 1+\frac{1}{2},\quad  1+\frac{1}{2},\quad \dots
\end{equation}
and the normalized eigenfunctions
\begin{equation}
\psi _n(x)=\frac{1}{\sqrt{n!\, 2^n\sqrt{\pi }}}\, {\rm e}^{-\frac{x^2}{2}}\, H_n(x).
\end{equation}
The general solution of the equation 
\begin{equation}
{\bf H}u=\varepsilon u
\end{equation}
considered up to a constant factor is
\begin{equation}
u_\varepsilon (x)={\rm e}^{-\frac{x^2}{2}}\left[ _1F_1\left( \frac{1}{4}-\frac{\varepsilon }{2};\frac{1}{2};x^2\right)
+2\mu x\frac{\Gamma \left(\frac{3}{4}-\frac{\varepsilon }{2}\right)}{\Gamma \left(\frac{1}{4}-\frac{\varepsilon }{2}\right)}
 {}_1F_1\left( \frac{3}{4}-\frac{\varepsilon }{2};\frac{3}{2};x^2\right) \right]
\end{equation}
where $\mu $ is an arbitrary constant \cite{sukum}.
For $\varepsilon <\frac{1}{2}$ and $|\mu |<1$ the solution $u_\varepsilon $ is nodeless and $1/u_\varepsilon $ is normalizable.
From the relation $H_0\, u_\varepsilon =\varepsilon \, u_\varepsilon $, that is,
\begin{equation}
-\frac{1}{2}\, u_\varepsilon ''+\frac{1}{2}x^2u_\varepsilon =\varepsilon \, u_\varepsilon 
\end{equation}
it follows the factorisation
\begin{equation}
{\bf H}-\varepsilon =A_\varepsilon ^+\, A_\varepsilon 
\end{equation}
where
\begin{equation}
A_\varepsilon =\frac{1}{\sqrt{2}}\left( -\frac{d}{dx}+\frac{u'_\varepsilon }{u_\varepsilon} \right),\qquad 
A^+_\varepsilon =\frac{1}{\sqrt{2}}\left( \frac{d}{dx}+\frac{u'_\varepsilon }{u_\varepsilon} \right).
\end{equation}
The supersymmetric partner 
\begin{equation}
{\bf H}_\varepsilon ={\bf H}-\frac{d^2\ln u_\varepsilon }{dx^2}=-\frac{1}{2}\, \frac{d^2}{dx^2}+\frac{1}{2}x^2-\frac{d^2\ln u_\varepsilon }{dx^2}
\end{equation} 
defined by the relation
\begin{equation}
{\bf H}_\varepsilon -\varepsilon =A_\varepsilon \, A_\varepsilon ^+
\end{equation}
has the eigenvalue spectrum 
\begin{equation}
\varepsilon,\quad  \frac{1}{2},\quad 1+\frac{1}{2},\quad  1+\frac{1}{2},\quad \dots
\end{equation}
and the corresponding normalized eigenfunctions 
\begin{equation}\fl
|0,\varepsilon \rangle \! \rangle \!=\!\frac{\frac{1}{u_\varepsilon}}{\sqrt{\int_{-\infty }^\infty \frac{1}{(u_\varepsilon (x))^2}dx}},\qquad 
|1,\varepsilon \rangle \! \rangle\!=\!\frac{A_\varepsilon \psi _0}{\sqrt{\frac{1}{2}-\varepsilon }},\qquad 
|2,\varepsilon \rangle \! \rangle\!=\!\frac{A_\varepsilon \psi _1}{\sqrt{1+\frac{1}{2}-\varepsilon }},\quad  \dots
\end{equation}
If we write the relation
\begin{equation}
\hat{H}\!=\! \frac{s\!+\!1}{2}\, |0;s\rangle\langle 0;s| + \sum_{n\geq 1}\, (n \!+\! s\!+\! 1/2)\, |n;s\rangle\langle n;s|
\end{equation}
as
\begin{equation}
\hat{H}-s-1\!=\! \frac{-s\!-\!1}{2}\, |0;s\rangle\langle 0;s| + \sum_{n\geq 1}\, (n \!-\! 1/2)\, |n;s\rangle\langle n;s|
\end{equation}
and choose 
\begin{equation}
\varepsilon =\frac{-s\!-\!1}{2},\qquad |0;s\rangle =|0,\frac{-s\!-\!1}{2} \rangle \!\rangle ,
\qquad |1;s\rangle =|1,\frac{-s\!-\!1}{2} \rangle \! \rangle , \dots
\end{equation}
then we get 
\begin{equation}
\hat{H}-s-1\!=\!{\bf H}_{\frac{-s\!-\!1}{2}}.
\end{equation}
On the other hand for 
\begin{equation}
A_{\vert z\vert^{2}, s} \,=\, \sum_{n=0}^{\infty}\, (n+ 2s + 1)\, |n;s\rangle\langle n;s|
\end{equation}
up to an isometry we have
\begin{equation}
A_{\vert z\vert^{2}, s} -2s-\frac{1}{2}={\bf H}
\end{equation}
where ${\bf H}$ is harmonic oscillator Hamiltonian. So, up to an isometry and a translation, the operator $\hat{H}$
is a supersymmetric partner of $A_{\vert z\vert^{2}, s}$.

\section{Some statistical properties}
\label{secstat}
\begin{figure}
\includegraphics{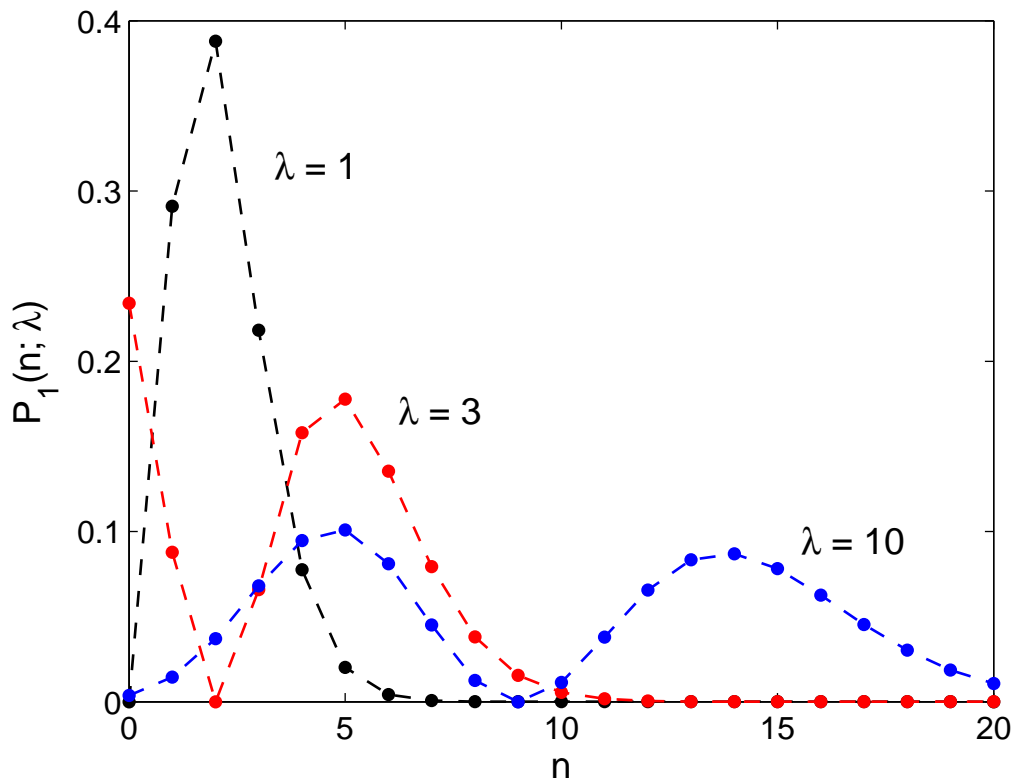}
\caption{\label{fig2} The distribution $P_{1}(n; \lambda)$ as given in eq. (\ref{eqS3}) is shown for $n$ varying from $0$ to $20$  by steps of $1$, and for $\lambda = 1$ (black dashed line), $3$ (red), and $10$ (blue). We note that the cancellation for $n = \lambda - 1$, and the existence of relative maxima at $n = 3; 0, 5; 5, 14$. }
\end{figure}

\begin{figure}
\includegraphics{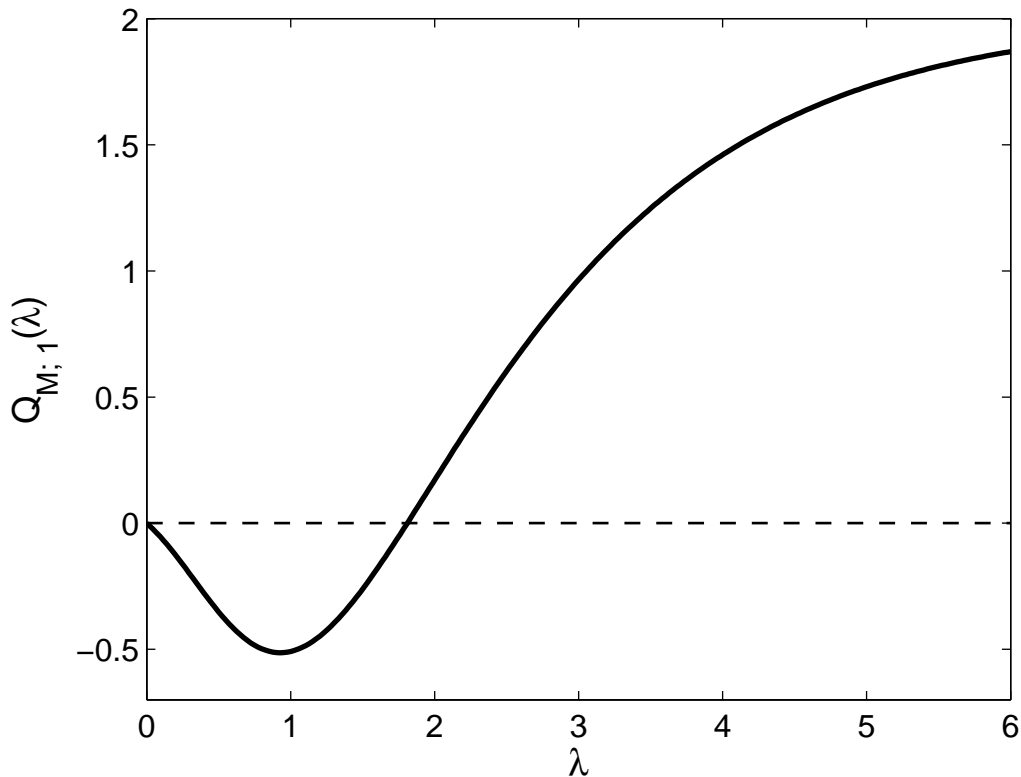}
\caption{\label{fig3} The so-called Mandel parameter $Q_{M; 1}(\lambda)$ as given by eq. (\ref{eqS4}). We see that $Q_{M; 1}(\lambda)$ is different from zero with the exception of one value of $\lambda=\lambda_{0}=1.81$. For $\lambda$ in $(0, \lambda_{0})$ (res. $(\lambda_{0}, \infty)$) we have a sub-Poissonian (res. super-Poissonian) distribution. For large values of $\lambda$ the parameter $Q_{M; 1}$ approaches to two.}
\end{figure}

Let us now examine some basic statistical aspects of the  coherent states (\ref{eqCS1}). Like the standard CS are  connected with the Poisson distribution, the complex Hermite CS are connected to the following generalization of the latter
\begin{equation}\label{eqS1} \fl 
n \mapsto P_{s}(n; \lambda) = \frac{1}{\mathcal{N}_s(\lambda)}\, 
\left(
\begin{array}{c}
s\!+\!n\\
s
\end{array}
\right)
\,\frac{\lambda^{n}}{n!}\, \left[_{1}F_{1}(-s; n+1, \lambda)\right]^{2} = \frac{s!}{\mathcal{N}_s(\lambda)}\, \frac{\lambda^{n}}{(s+n)!}\,\left(L_s^{(n)}(\lambda) \right)^2\, .
\end{equation}
The parameter $\lambda \in\mathbb{R}$ is equal to $\vert z\vert^2$. For $s=0$ the distribution (\ref{eqS1}) reduces to the Poisson distribution with parameter $\lambda$. 
For $s \neq 0$, a quantitative estimate of the deviation from Poisson statistics is provided by  the so-called Mandel parameter $Q_{M} \deq (\Delta n)^2/\langle n\rangle -1$, where $(\Delta n)^2 = \langle n^2 \rangle - \langle n\rangle^{2}$ is a variance calculated for a given distribution. It is well-known that in the Poissonian case we have $Q_{M} = 0$ while for $Q_{M} < 0$ (resp. $Q_{M} >0$) we say that the distribution is sub-Poissonian (resp. super-Poissonian). %The Mandel 
Without loss of generality let us consider the probability distribution and the Mandel parameter for $s=1$. In this case from eq. (\ref{eqS1}), we get the following expression for the distribution
\begin{equation}\label{eqS3}
P_{1}(n; \lambda) = e^{-\lambda}\,(\lambda^{n}/n!)\,\frac{n+1}{1 - e^{-\lambda}\,\lambda}\,\left(1 - \frac{\lambda}{n+1}\right)^2\, ,
\end{equation}
where we can identify the corrective factor to the Poisson distribution, 
and the for following Mandel parameter,
\begin{equation}\label{eqS4}
Q_{M; 1}(\lambda) = -{\frac {{e^{\lambda}}{\lambda}^{2}+2\,{e^{\lambda}}+4\,{e^{\lambda}}\lambda-2\,{e^{2\,\lambda}}-\lambda}{ \left({e^{\lambda}}-\lambda \right)  \left( 1+{e^{\lambda}} \right) }}
%\frac{\lambda(2e^{\lambda} - \lambda e^{\lambda} - 1)}{(e^{\lambda} - 1)(e^{\lambda} - \lambda)}, \quad {\rm for}\quad \lambda\neq 0.
\end{equation}
The behavior of the distribution $P_{1}(n; \lambda)$ for three values of the  parameter $\lambda$, namely $1$, $3$, and $10$, is shown in Fig. \ref{fig2}. The behavior of the parameter $Q_{M; 1}$ is shown in Fig. \ref{fig3}. There we note the subpoissonian character of the distribution for $\lambda   < 1.81$. The latter becomes superpoissonian for $\lambda   > 1.81$ while going smoothy to zero as $\lambda$ becomes large. 
%Parameter $Q_{M; 1}$ is shown  in Fig. (\ref{fig3}).

\section{Hermite Quantization of the real line}
\label{secrealmerm}

\begin{figure}
\includegraphics[scale=0.4]{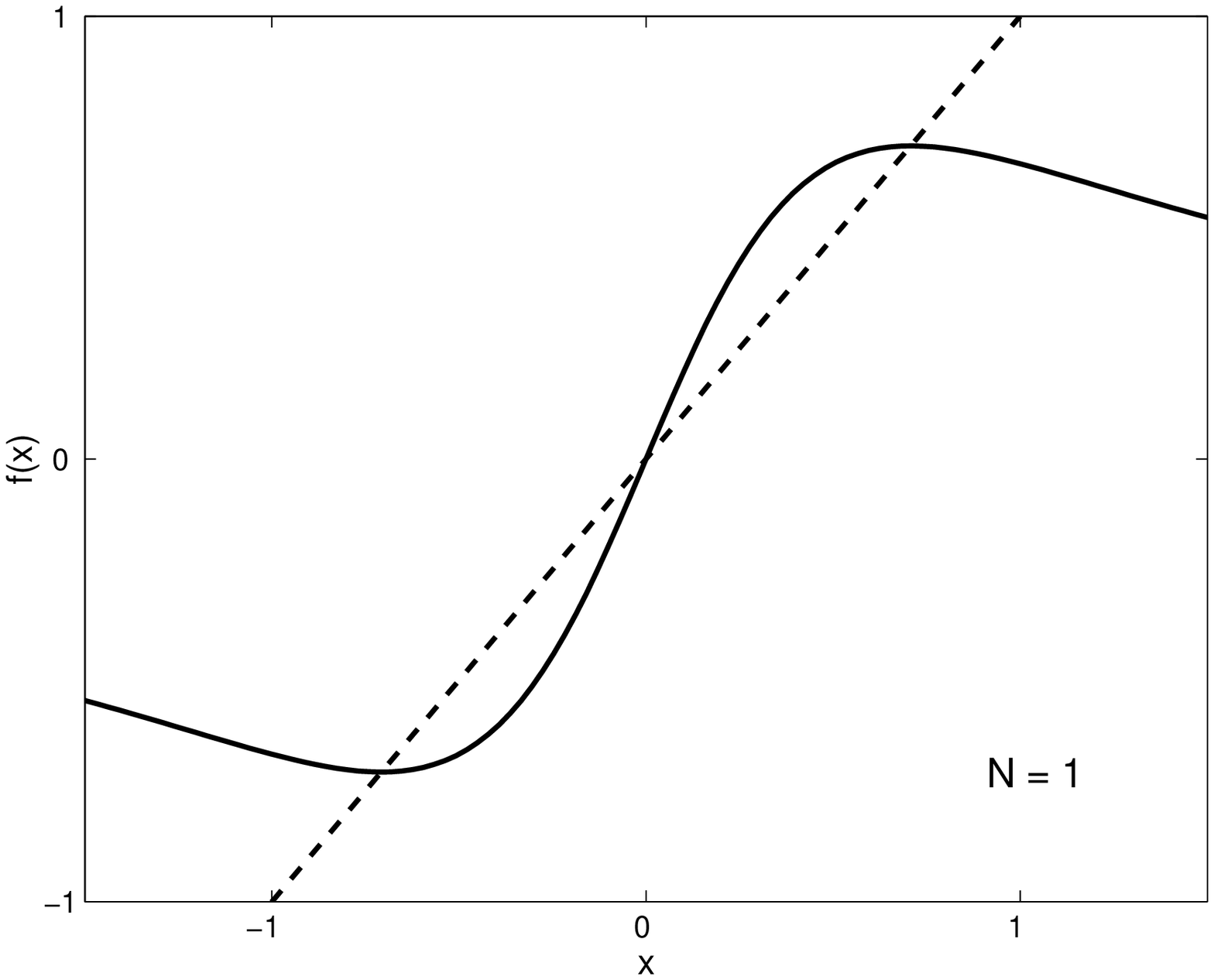}

\includegraphics[scale=0.4]{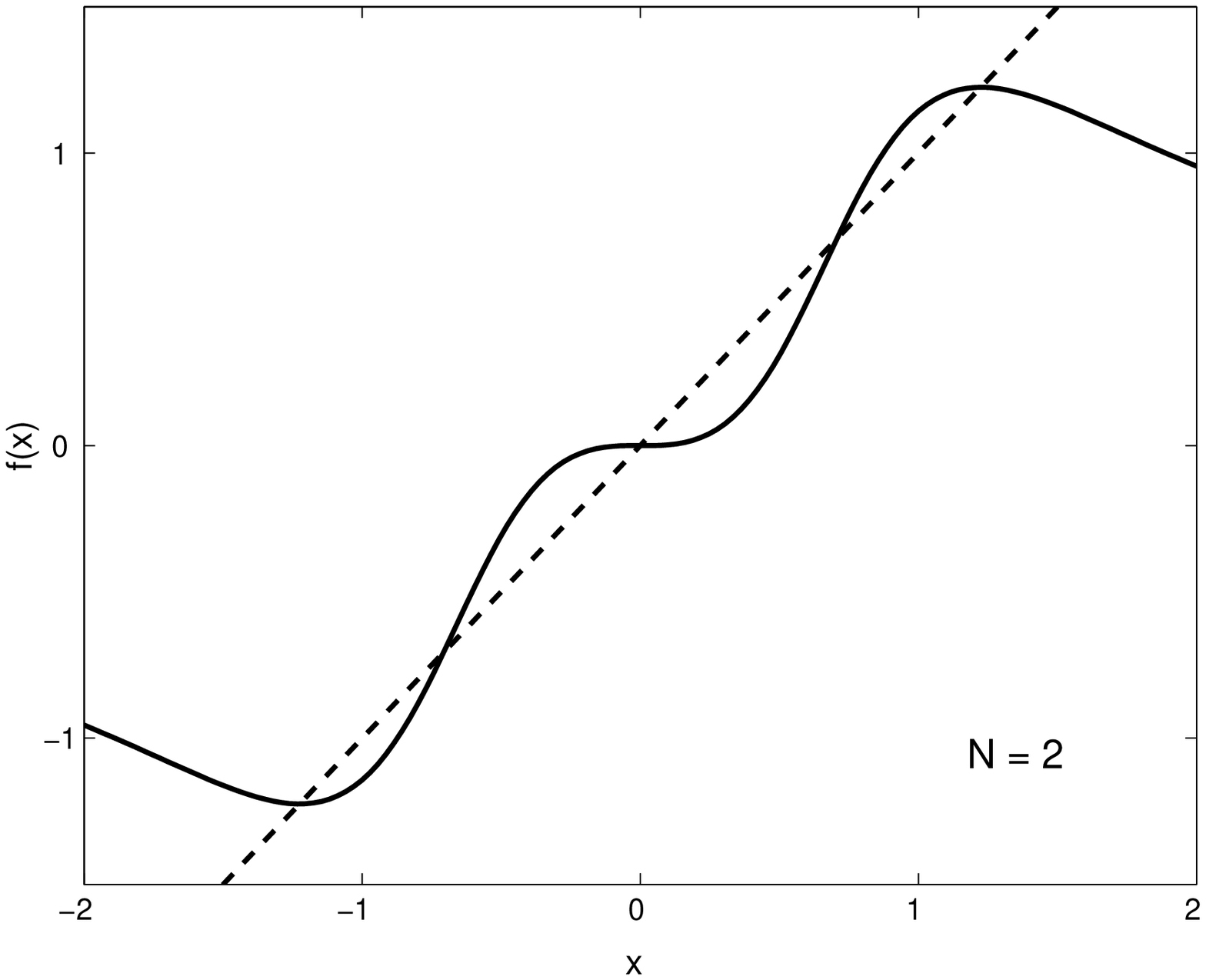}

\includegraphics[scale=0.4]{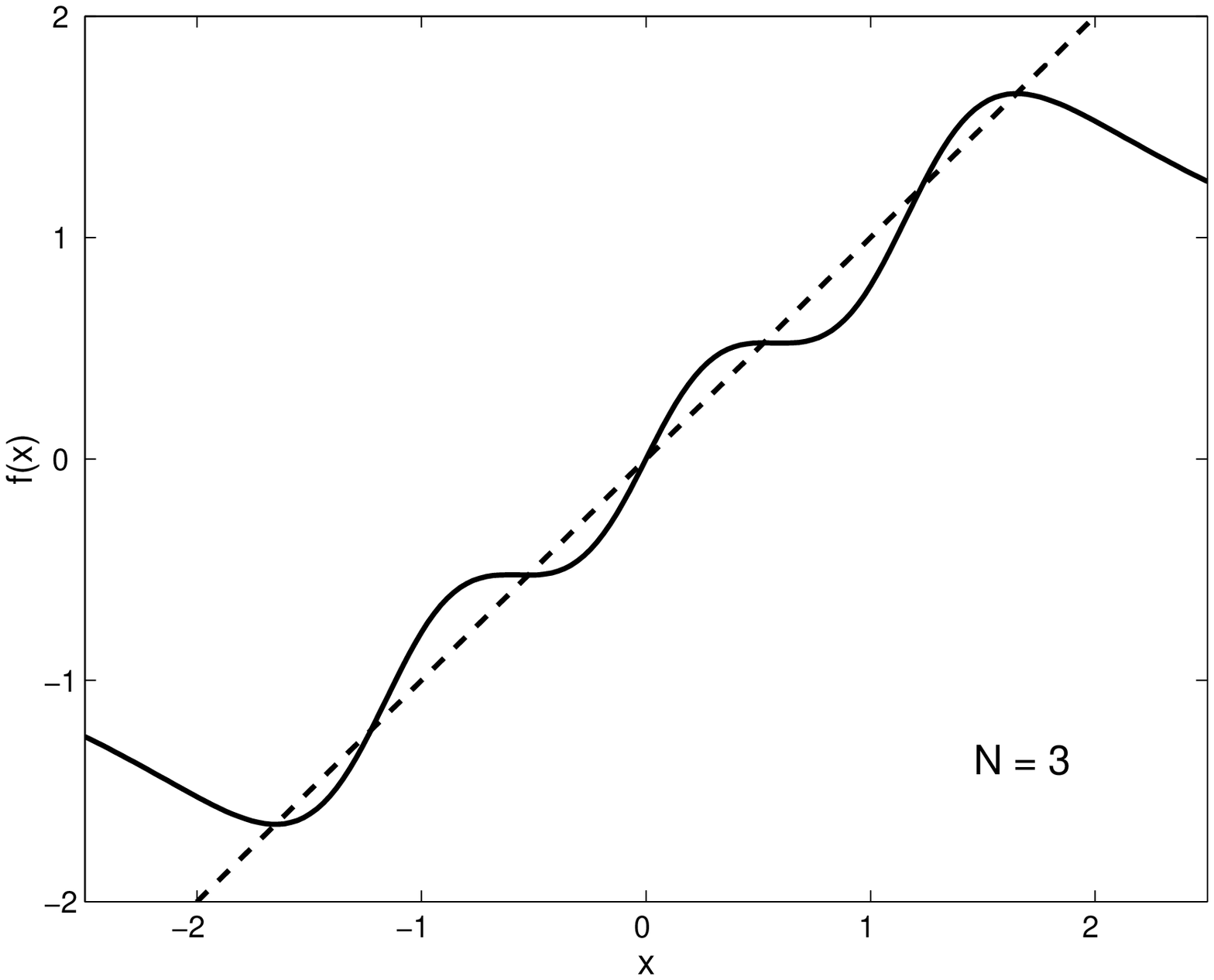}

\includegraphics[scale=0.4]{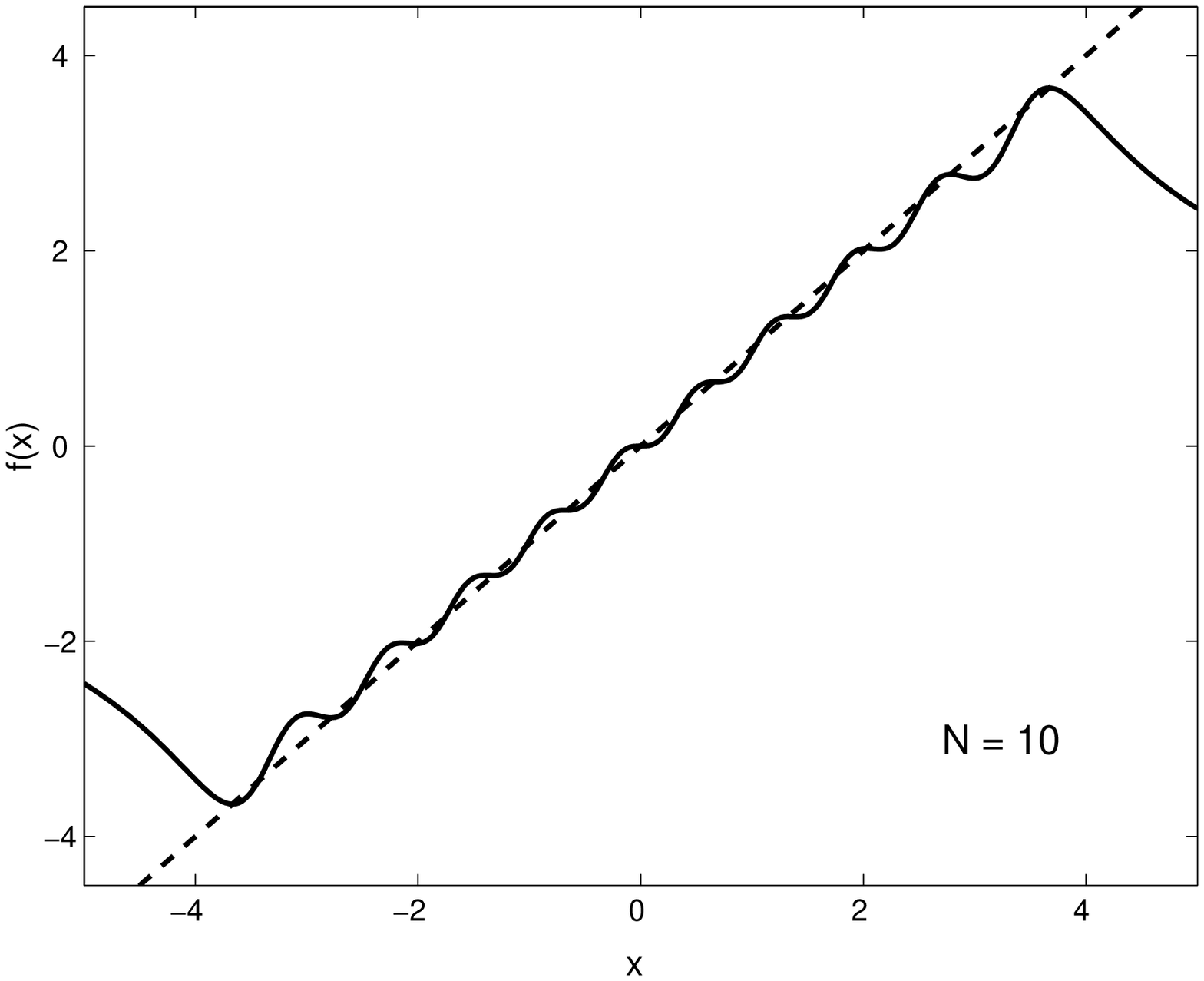}
\caption{\label{fig1} The lower symbol (full line) as given in eq. (\ref{eqHQ7}) for $f(x) = x^{r}$ and $r=1$, $N = 1, 2, 3$ and $10$. The dashed line is the classical quantity $f(x) = x$.}
\end{figure}

Since we are examining in this paper some aspects of complex Hermite polynomials related to quantization, it is  interesting to explore as well the same aspects for \emph{real} Hermite polynomials. 
It is well-known that Hermite polynomials $H_0$, $H_1$, $H_2$, \dots  form an orthogonal basis of the Hilbert space $L^2(\R,\, dx)$. 
\begin{equation}\label{eqRH1}
\frac{1}{\sqrt{\pi }}\int _{-\infty }^\infty {\rm e}^{-x^2}H_m(x)\, H_n(x)\, dx=
\left\{ \begin{array}{lll}
n!\, 2^n & {\rm if} & m=n\\[2mm]
0 & {\rm if} & m\not=n
\end{array} \right.
\end{equation}
Now, $L^2(\R,\, dx)$ is \underline{not} a reproducing kernel Hilbert space, a required property for building coherent states  
resolving the unity \cite{G2000}. This reflects in the fact that $\sum_{n=0}^{\infty} (H_n(x))^2 = \infty$. The most we can do here is to deal with finite subsets of such polynomials. Since they 
satisfy Christoffel-Darboux formula
\begin{equation}\label{eqRH2}
\sum_{n=0}^{N}\frac{1}{n!\, 2^n}H_n(x)\, H_n(y)=\frac{H_{N+1}(x)\, H_{N}(y)-H_{N}(x)\, H_{N+1}(y)}{N!\, 2^{N+1}\, (x-y)}
\end{equation}
and its direct consequence
\begin{equation}\label{eqRH3}
\sum_{n=0}^N \frac{1}{n!\, 2^n}H_n^2(x) = \left[H_{N+1}^2(x)-H_N(x)\,H_{N+2}(x)\right]/(N!\, 2^{N+1})\, ,
\end{equation}
let us take the most of this formula in exploring ``real Hermite'' quantization of the real line. 
Let $\mathcal{E}_N$ be a real Hilbert space and $\{ e_0,\, e_1,\, \dots \, , \, e_N\}$ an orthonormal basis
in $\mathcal{E}_N$. The system of unit vectors
\begin{equation}\label{eqHQ1}
|x\rangle =\frac{1}{\sqrt{\mathcal{N}_{N}(x)}}\sum_{n=0}^N\frac{1}{\sqrt{n!\, 2^n}} H_n(x)\, |e_n\rangle , \qquad x\in \mathbb{R}
\end{equation}
with 
\begin{equation}\label{eqHQ2}
\mathcal{N}_{N}(x)=\sum_{n=0}^N\frac{1}{n!\, 2^n}H_n^2(x)=\frac{H_{N+1}^2(x)-H_N(x)\, H_{N+2}(x)}{N!\, 2^{N+1}}
\end{equation}
satisfy the resolution of identity 
\begin{equation}\label{eqHQ3}
\frac{1}{\sqrt{\pi }}\int _{-\infty }^\infty {\rm e}^{-x^2}\, \mathcal{N}_{N}(x)\,|x\rangle \langle x| dx
=\mathbb{I}_{\mathcal{E}_N}
\end{equation}
and the overlapping relation
\begin{eqnarray}\label{eqHQ4}
\langle x|y\rangle &=& \frac{1}{\sqrt{\mathcal{N}_{N}(x)\, \mathcal{N}_{N}(y)}}\, \sum_{n=0}^N \, \frac{1}{n!\, 2^{n}}\, H_n(x)\, H_n(y)\\[5mm]
&=& \frac{H_{N+1}(x)H_{N}(y)\, - \,H_{N}(x)H_{N+1}(y)}{N!\, 2^{N+1}(x-y)\, [\mathcal{N}_{N}(x)\, \mathcal{N}_{N}(y)]^{1/2}} 
\end{eqnarray}
The system $\{ |x\rangle \}_{x\in \mathbb{R}}$ is a continuous frame in $\mathcal{E}_N$. It allows us to
associate to each function $f:\mathbb{R}\longrightarrow \mathbb{R}$ 
satisfying certain conditions a linear operator, namely,
\begin{equation}\label{eqHQ5}
A_f:\mathcal{E}_N\longrightarrow \mathcal{E}_N, \qquad 
A_f=\frac{1}{\sqrt{\pi }}\int _{-\infty }^\infty |x\rangle \, {\rm e}^{-x^2}\, \mathcal{N}_{N}(x)\,f(x)\,dx \,  \langle x|.
\end{equation}
The lower symbol $\check f:\mathbb{R}\longrightarrow \mathbb{R}$,
\begin{equation}\label{eqHQ6}
\check f(t)=\langle t|A_f|t\rangle =
\frac{1}{\sqrt{\pi }}\int _{-\infty }^\infty {\rm e}^{-x^2}\, \mathcal{N}_{N}(x)\,f(x)\, |\langle x|t\rangle |^2 dx
\end{equation}
is given by
\begin{equation}\label{eqHQ7} \fl
\check f(t) \!=\! \frac{1}{(N!)^2 4^{N+1}\sqrt{\pi}\mathcal{N}_{N}(t)}\int_{-\infty }^\infty 
\frac{{\rm e}^{-x^2}}{(t-x)^2}f(x)\, [H_{N+1}(t)H_N(x)\!-\!H_N(t)\, H_{N+1}(x)]^2 dx.
\end{equation}
The integrals (\ref{eqHQ7}) can be easily calculated for $N= 0$ and $f(x) = x^{r}$ ($r\in\mathbb{N}$). The lower symbol vanishes for all odd $r = 2k_{1}+1$ ($k_{1}\in\mathbb{N}$), whereas is equal to $1$ or $(2k_{2}-1)!!/\,2^{k_{2}}$, respectively for $r = 0$ or $r = 2k_{2}$ ($k_{2} = 1, 2, \cdots$). For odd $r$ and for $N\neq 0$ the first non-zero value of $\check{f}$ in eq. (\ref{eqHQ7}) is for $r=1$ and for $N = 1$, it is $2t/(1+2t^2)$. The behavior of lower symbols for $r=1$ and for $N = 1, 2, 3$ and $10$ is shown in Fig. \ref{fig1}. The dashed lines are the classical quantity, $f(x) = x$, while the full line denotes $\check f(t)$. We can see that the graph of $\check f(t)$ wraps its classical counterparts only in median sector, which enlarges with the increasing $N$.  
 
Now, let us calculate the operator $A_{f}$ for classical quantities defined as infinite sum of Hermite polynomials, 
$f(x) = \sum_{n=0}^{\infty} a_{n} H_{n}(x)$, where $a_{n} = \frac{1}{n!\, 2^{n}}\langle H_{n}| f\rangle $.
For such a choice of $f$, with the help of equation (\ref{eqHQ5}) and the formula 2.20.17.2 in \cite{Prud1} the operator $A_{f}$ reads as
\begin{equation}\label{eqHQ8}
A_{f} = \sum_{k,l=0}^{N}\,A_{f}^{k, l}\,|k\rg\lg l|\, , 
\end{equation}  
where
\begin{equation}\label{eqHQ9}
A_{f}^{k, l} = \sum_{r=0}^{\min(k,l)}\, a_{k+l-2r}\, 2^{(k+l-2r)/2}\, \frac{(k+l-2r)!\,\sqrt{k!l!}}{(k-r)!\,(l-r)!\,r!}.
\end{equation}
For fixed value of the parameter $N$ the operator $A_{f}$ depends only on a few first coefficients $a_{k}$. It means that an infinite set of classical quantities leads to the same operator and we lose a lot of information about classical systems.

The position operator $A_{x}$ can be calculated by using the equations (\ref{eqHQ8}), (\ref{eqHQ9}) and the well-known relations
\begin{equation}\label{eqHQ10}
x^{r} = (r!/2^{r})\,\sum_{k=0}^{[r/2]} H_{r-2k}(x)/\left[k!\, (r-2k)!\right]
\end{equation}
The symbol $[\cdot]$ denotes the integer part of the involved number. The equation (\ref{eqHQ10}) for $r = 1$ leads to $x = (1/2)H_{1}(x)$. Thereby, in eq. (\ref{eqHQ9}), only the coefficient $a_{1}$ and the terms with $k, k+1$ or $k+1, k$ are different from zero. The explicit form of the operator $A_{x}$ is given as the $(N+1)\times(N+1)$ matrix  
\begin{equation}\label{eqHQ11}
A_{x}= \left(
\begin{array}{lllllll}
	0 & \frac{1}{\sqrt{2}} & 0 & \cdots & 0\\
	\frac{1}{\sqrt{2}} & 0 & 1 & \cdots & 0\\
	0 & 1 & 0 & \ddots & \vdots \\
	\vdots &  & \ddots & \ddots & \sqrt{\frac{N}{2}} \\
	0 & \cdots & 0 & \sqrt{\frac{N}{2}} & 0\\
\end{array}\right)
\end{equation}

We note that (\ref{eqHQ11}) is the same as the finite approximation $Q_{N}$, of the position operator $Q$ in usual Quantum Mechanics. This $Q_{N}$ is obtained from $q = (z +\bar{z})/\sqrt{2}$ by quantization with finite approximation of standard coherent states \cite{gajomon}. 

The spectral properties of the position operator $A_{x}$ are the same as for $Q_{N}$ in \cite{gajomon}. The characteristic equation $\Lambda_{N} = \det\left(A^{N}_{f} - \lambda\mathbb{I}_{N}\right)$, satisfies the following relation
\begin{equation}\label{eqHQ16} 
\Lambda_{N+1}(\lambda) = \lambda\Lambda_{N} + (N/2)\,\Lambda_{N-1}(\lambda),
\end{equation}
which, for $\Lambda_{N} = (-2)^{-N}H_{N}(\lambda)$, leads to the recurrence relation for the Hermite polynomials $H_{N}(\lambda)$.

\section{Conclusion}
\label{seconc}
We have explored some unexpected features of coherent state quantization of the complex plane and of the real line  and of some functions living on them. The complex plane can be viewed as the phase space for the motion of a particle on the real line, and we have shown that there exist infinitely many ways to analyze it from a quantum perspective. The fundamental question that can now be addressed from our results is the existence or not of an actual ``canonical'' or ``privileged'' point of view among that infinite set of possibilities, uniquely discriminated on experimental bases. The answer goes far beyond the scope of this paper. Concerning our  ``quantum version'' of the real line, we have shown that coherent state quantization  yields localization properties quite similar to those revealed by ordinary quantum mechanics. One is naturally led to conclude from these rather elementary facts that the (long!) quest for a univocal quantum version of a ``classical'' object may reveal unexpected surprises, and open the way to a large field of future investigations.

\section*{Acknowledgments}
The authors thank S.T. Ali and F. Bagarello  for helpful conversations on the subject matter of this paper. NC acknowledges the support provided by CNCSIS under the grant IDEI 992 - 31/2007.

..................

\end{document}